\documentclass[12pt]{article}
\usepackage{graphicx}

\begin{document}

\noindent
{\large Relativistic vs. Non-relativistic Nuclear Models}

\vspace{1cm}
\noindent
Haruki KURASAWA\\
Department of Physics, Faculty of Science, Chiba University,\\
Chiba 263-8522, Japan\\
Tel +81-43-290-3687\\
Fax +81-43-290-3691\\
kurasawa@c.chiba-u.ac.jp\\

\vspace{2mm}
\noindent
Toshio SUZUKI\\
Department of Applied Physics, Fukui University, Fukui 910-8507, Japan\\
and\\
RIKEN, 2-1 Hirosawa, Wako-shi, Saitama 
351-0198, Japan\\
Tel +81-776-27-8780\\
Fax +81-776-27-8494\\
suzuki@quantum.apphy.fukui-u.ac.jp

\vspace{5mm}
\noindent
Abstract

Both the relativistic and non-relativistic model
explain very well low-energy nuclear phenomena, but 
in a physically different way from each other.
There seems to be no low-energy phenomenon to answer which model
is more reasonable. 
In order to explore a difference between two models,
high momentum transfer phenomena are investigated.
First it is shown that the neutron spin-orbit charge density 
in the relativistic model reproduces very well experimental data
on elastic electron scattering which have not been explained in 
the non-relativistic model. Next it is  predicted that
the relativistic Coulomb sum value is strongly quenched,
compared with the non-relativistic one. This quenching is
owing to the anti-nucleon degrees of freedom,
which make the nucleon size larger and equivalently the 
vector-meson mass smaller in nuclear medium. 
New experiment on the Coulomb sum values around the momentum transfer 
1GeV is expected to distinguish the relativistic from the 
non-relativistic model.

\vspace{2mm}
\noindent
[relativistic models vs. non-relativistic models, nuclear structure,
electron scattering, Coulomb sum rule]


\newpage

\section{Introduction}

There are two kinds of nuclear models which have been extensively
studied in recent years. The one is a non-relativistic model
which assumes Skyrme forces\cite{c}, and the other is
a relativistic model which takes into account
meson exchanges explicitly\cite{se}. Both models work very well
phenomenologically in reproducing the nuclear ground state and
low-lying excited states 
including giant resonances. 
The problem, however, is that the relationship is not clear
between the non-relativistic and relativistic model. 
The former cannot be obtained in a simple non-relativistic reduction
of the latter,
or the latter is not a simple relativistic extension of the former.
In the relativistic model the effective nucleon-mass coming from
the Lorentz-scalar potential and anti-nucleon degrees of freedom
play an essential role even in low-energy phenomena, while in the
non-relativistic model many-body correlations are necessary.
These models now understand nuclear structure in different ways from
each other, as briefly reviewed in the next section.

At present
there seems to be no low-energy phenomenon to distinguish the one
from the other model.
In this paper, however, we would like to point out that the 
relativistic effects hidden in low-energy phenomena 
manifest themselves in high-energy phenomena in a way 
peculiar to the relativistic model. 
Since, if the relativistic model is realistic, high-energy phenomena
should be also described within the same framework, such
phenomena have a possibility to distinguish
the relativistic model from the non-relativistic model.

We will show in the third section 
that the relativistic model provides us with a peculiar 
time-component of the nuclear four-current. The neutron spin-orbit 
charge 
density is enhanced by the Lorentz scalar potential. 
As a result, the difference between cross sections for 
$^{40}$Ca and $^{48}$Ca is well reproduced, which has not been
 explained for long time in non-relativistic models. 

In the fourth section, we will show 
how effects of the anti-nucleon
degrees of freedom appear in 
high-momentum transfer reaction.
In the relativistic model, the anti-nucleon degrees of freedom 
are necessary for describing correctly the convection current, 
the center of mass motion, and giant
resonances, but they do not change non-relativistic results at all. 
In the same framework as for those low-momentum 
transfer phenomena, we will show that the Coulomb sum values
at the high-momentum transfer are strongly 
quenched owing to the anti-nucleon degrees of freedom. 
This quenching is 
due to the fact that in the relativistic model, the nucleon size is 
increased and equivalently the vector-meson mass is decreased 
owing to the presence of the Lorentz scalar potential
in nuclear medium. 

The final section will be devoted to a brief conclusion.

\section{Relativistic vs. Non-relativistic Model}

In this section, we briefly review how the 
relativistic and non-relativistic model explain
low-energy nuclear phenomena in different ways from
each other. As examples, let us quote the binding energy
and giant monopole states.

\subsection{Binding Energy}
The one of the Skyrme forces in the non-relativistic model yields
the the nucleon binding energy in nuclear matter as\cite{c}
\begin{equation}
 \frac{E_{\rm N}}{A} = \frac{k_{\rm F}^2}{2M} + \left(\frac{3}{4}
+ \frac{3}{16}t_3\rho\right) + \frac{1}{10}\left(3t_1 +
5t_2\right)\rho k_{\rm F}^2,
\end{equation}
where $k_{\rm F}$ and $\rho$ denote the Fermi momentum and
the nucleon density, and $t_0, t_1, t_2$ and $t_3$ stand for the
parameters in the Skyrme force. The repulsive $t_3$-force is
necessary for the non-relativistic model in order to prevent
the nucleus from the collapse, and is considered to simulate
many-body correlations.
On the other hand, the simple relativistic model gives the
binding energy as\cite{se}
\begin{equation}
\frac{E_{\rm R}}{A} = \frac{1}{\rho}\left\{\frac{1}{(2\pi)^3}
\int_0^{k_{\rm F}}dk^3(\vec{k}^2 + M^{*2})^{1/2} +
\frac{1}{2}\left(\frac{m_{\rm s}}{g_{\rm s}}\right)^2
U_{\rm s}^2
+ \frac{1}{2}\left(\frac{m_{\rm v}}{g_{\rm v}}\right)^2
U_0^2 \right\}, \label{en}
\end{equation}
where $U_{\rm s}$ and $U_0$ represent the Lorentz-scalar and -vector
potential given by the Lorentz-scalar
and -vector density as
\begin{eqnarray}
U_{\rm s} = \left(\frac{g_{\rm s}}{m_{\rm s}}\right)^2
\rho_{\rm s}, \ \ \ \ \
U_{\rm v} = \left(\frac{g_{\rm v}}{m_{\rm v}}\right)^2
\rho,
\end{eqnarray}
$m_{{\rm s(v)}}$ and $g_{{\rm s(v)
}}$ being the
Lorentz-scalar(vector)
meson mass and Yukawa coupling constant.
The effective mass in Eq.(\ref{en}) comes from the
Lorentz-scalar potential,
\begin{eqnarray}
 M^* = M - U_{\rm s}.
\end{eqnarray}
As seen in Eq.(\ref{en}), the higher-order density dependence
of the binding energy comes from the relativistic effects,
in contrast to the one in the non-relativistic model. The Lorentz
scalar potential which yields the effective mass prevents
the nucleus from the collapse.

\subsection{Giant Monopole States}

An important ingredient of the relativistic model is anti-nucleon 
degrees of freedom. They are necessary even for low-energy
phenomena. For example, the excitation energy of the giant monopole state
is described in terms of the Landau parameters as 
in non-relativistic models\cite{ni},
\begin{eqnarray}
\omega_0 = \frac{1}{\epsilon_F}\left\{\frac{3k_F^2}{\langle r^2 \rangle}
\frac{1+F_0}{1+\frac{1}{3}F_1}\right\}^{1/2},\label{mo}
\end{eqnarray}
where $F_0$ and $F_1$ denote the Landau parameters, and $\epsilon_F$
the Fermi energy. In the relativistic model, the anti-nucleon 
degrees of freedom are hidden in the Landau parameters. 
In the $\sigma-\omega$ model, they are described as\cite{k2} 
\begin{eqnarray}
F_0 = F_{\rm v} - \frac{1-v_F^2}{1+a}F_{\rm s}, \ \ \ \ 
F_1 = - \frac{v_F^2F_{\rm v}}{1+\frac{1}{3}v_F^2F_{\rm v}}, \label{lp}
\end{eqnarray}
where we have defined
\begin{eqnarray}
F_{\rm s} &=& N_F(\frac{g_{\rm s}}{m_{\rm s}})^2, \ \ F_{\rm v} 
= N_F(\frac{g_{\rm v}}{m_{\rm v}})^2, \ \ 
N_F = \frac{2k_FE_F}{\pi^2}, \ \ E_F = \{k_F^2 + M^{*2}\}^{1/2},
\nonumber\\  
v_F &=& k_F/E_F, \ \ 
a = \frac{3}{2}( 1- \frac{2}{3}v_F^2
 + \frac{1-v_F^2}{2v_F}\ln \frac{1-v_F}{1+v_F}).\nonumber
\end{eqnarray}
It has 
been shown that the denominators of Eq.(\ref{lp}) come from
nucleon-antinucleon excitations in the configuration space of RPA\cite{k3}.
The similar role of the anti-nucleon in the relativistic model 
can be found in excitation energies of other giant resonances and
the nuclear convection current or magnetic moments\cite{k4,b}. In particular,
the continuity equation is not satisfied without nucleon-antinucleon 
excitations\cite{k3}.

Thus, in the relativistic model, the anti-nucleon degrees
of freedom are necessary even for low-energy phenomena,
but we cannot distinguish it from the non-relativistic
model by these phenomena, since there is no difference
between their relativistic and non-relativistic expressions,
as in Eq(\ref{mo}).

We note that in Eq.(\ref{lp}) the relativistic effects
are also important. If the effects of $O(v^2_F)$ are neglected,
we have 
$F_0 < - 1$ which implies the nuclear collapse. Furthermore,
we have $F_1 = 0$. It comes from
the space part of the $\omega$-meson exchange which is
usually neglected in the non-relativistic model.

\section{Elastic Electron Scattering}

Now let us discuss high momentum transfer phenomena 
where relativistic effects are expected to appear.
Of course non-relativistic models do not contain
such effects. Hence, we must find
relativistic effects which are not a simple correction
to non-relativistic models, but are peculiar to the 
present relativistic model.

First we discuss elastic electron scattering from nuclei.
In order to calculate the cross section 
in phase-shift analyses, we need
the nuclear charge density,
\begin{eqnarray}
\rho_c(r) = \int\frac{d^3q}{(2\pi)^3}\exp(-i\vec{q}\cdot\vec{r})
 \langle 0|\hat{\rho}(\vec{q})|0\rangle\label{c}
\end{eqnarray}
where the time-component of the relativistic nuclear-four current
is given by
\begin{eqnarray}
\langle 0|\hat{\rho}(\vec{q})|0\rangle
=\langle 0| \sum_k\exp(i\vec{q}\cdot\vec{r}_k) \left(
F_{1k}(\vec{q}^{\,2})
+\frac{\mu_k}{2M}F_{2k}(\vec{q}^{\,2})\vec{q}\cdot\vec{\gamma}_k
\right)|0\rangle .
\end{eqnarray}
Using the Sachs form factor instead of the Dirac form
factor $F_1(\vec{q}^{\,2})$,
it is rewritten as
\begin{eqnarray}
\langle 0|\hat{\rho}(\vec{q})|0\rangle
= \int d^3x\exp(i\vec{q}\cdot\vec{x})
\sum_\tau\left(G_{E\tau}(\vec{q}^{\,2})\rho_\tau(x)
	  +F_{2\tau}(\vec{q}^{\,2})W_\tau (x)\right),
\end{eqnarray}
where $\tau$ denotes the proton and neutron. 
The nucleon density $\rho_\tau(r)$
and the spin-orbit density $W_\tau(r)$ in the above equation
are given in the relativistic model as
\begin{eqnarray}
\rho_\tau(r) &=& \sum_{\alpha}\frac{2j_\alpha+1}{4\pi r^2}
\left( G_\alpha^2+F_\alpha^2 \right),\\
W_\tau(r) &=& \frac{\mu_\tau}{M}\sum_\alpha
\frac{2j_\alpha+1}{4\pi r^2}\nonumber\\
& & \times \frac{d}{dr}\left( \frac{M-M^\ast(r)}{M}\,G_\alpha F_\alpha
+\,\frac{\kappa_\alpha+1}{2Mr}\,G_\alpha^2-
\frac{\kappa_\alpha-1}{2Mr}\,F_\alpha^2\right), \label{rsd}
\end{eqnarray}
where
\begin{eqnarray}
\mu_p = 1.793, \ \ \mu_n = -1.913, \ \ 
\kappa_\alpha = (-1)^{j-\ell+1/2}(j+1/2).
\nonumber 
\end{eqnarray}
In the non-relativistic model, we have the only large
component $G_\alpha$, but no small component $F_\alpha$
of the single-particle wave function.
As a result, there is no spin-orbit density in the non-relativistic
model, since it is in order of $1/M^2$, while
Schr$\ddot{{\rm o}}$dinger
equation is in order of $1/M$. In the relativistic model, on the
other hand, we have the spin-orbit density coming 
from both protons and neutrons. In the present relativistic model,
it is enhanced by the
effective mass. The enhancement is more clearly seen in
the Fourier transform of Eq.(\ref{rsd}):
\begin{eqnarray}
 W_{\tau}(q)\approx \frac{\mu_\tau}{M}q\sum_\alpha (2j_\alpha+1)
\int_0^\infty
dr\frac{\kappa_\alpha+1}{2M_*r}G_\alpha^2j_1(qr),
\end{eqnarray}
where we have defined
\begin{eqnarray}
2M_*(r)=E+M^*(r)-U_0(r)\approx 2M^*(r).
\end{eqnarray}
This spin-orbit density is not negligible, in particlar,
in neutron-rich
nuclei where the subshell is not occupied by the protons.

An example\cite{k1} is shown in Fig. 1 which shows 
the elastic-scattering cross sections for 
$^{40}$Ca and $^{48}$Ca on the top and their difference $D(\theta)$
on the bottom as a function of the electron-scattering angle in the 
case of the incident energy 249.5 MeV. The solid curves on the l.h.s.
 of Fig.1 are obtained
in the non-relativistic model using the SLy4 force\cite{c}. We clearly see 
a disagreement with experiment\cite{f} on the difference between
the two cross
sections. Other Skyrme forces available at present yield similar
results for the difference. We note that the neutron charge densities 
are taken into account properly in these non-relativistic calculations,
although their contribution to the cross sections are negligible.
On the other hand, the solid curves
of the r.h.s. of Fig. 1 are
obtained in the relativistic model with NL-SH parameter set\cite{sh}.
The thick and thin solid curves are calculated with and without 
the spin-orbit density, respectively. 
In $^{40}$Ca, the two curves coincide with each other,
since the spin-orbit
density has almost no effects in doubly closed shell
nuclei. It is seen that the 
difference between the cross sections for $^{40}$Ca and  $^{48}$Ca
 is well reproduced in the
relativistic model owing to the neutron spin-orbit density
peculiar to the relativistic model. 
If electron scattering off unstable nuclei becomes available in the
near future, we expect that
the difference  between the relativistic and non-relativistic models
would be explored in more detail.


\section{Quasielastic Scattering}


In the low-energy phenomena mentioned in the subsection 2.2,
we cannot see effects of
the anti-nucleon degrees of freedom explicitly, since
they do not change
the results of the non-relativistic model at all.
We notice, however,
that those phenomena are all related to the zero limit of the 
momentum transfer $q$ in the RPA response functions\cite{k3}. 
At the finite momentum transfer, effects of the nucleon-antinucleon
excitations
appear in the RPA response function. This fact has been shown in the
longitudinal response function for quasielastic scattering\cite{k5,k6}. The 
response function is strongly quenched by the effects of
order $\vec{q}^{\,2}$ as discussed below.

Fig. 2 shows the Coulomb sum values as function of the momentum transfer.
The dashed and solid curves show the results of the relativistic RPA 
without and with the nucleon-antinucleon excitations, respectively.
The non-relativistic results are similar to the dashed one.
It is seen that above the momentum transfer 0.5 GeV, the Coulomb sum
values are strongly quenched due to the nucleon-antinucleon
excitations.


The reason why the Coulomb sum values are quenched is easily
understood in Fig. 3. 
The Dirac form factor $F_1$ in the free space includes the second
diagram of the r.h.s. in which the nucleon-antinucleon excitations participate 
with the free mass $M$. In nuclear medium, however, nucleons and
anti-nucleons have the effective mass $M^* = M - U_{\rm s}$ in the 
relativistic model. This effective mass modifies the Dirac 
form factor, that is the proton radius. Using the RPA correlation
function, the modification of the proton size is estimated as\cite{k7}
\begin{eqnarray}
\langle r_p^2 \rangle^* = \langle r_p^2 \rangle + \delta \langle r_p^2
\rangle, \ \ \ 
\delta \langle r_p^2 \rangle \approx \frac{1}{\pi^2}
\left(\frac{g_{\rm v}}{m_{\rm v}}\right)^2\ln \left(\frac{M}{M^*}\right).
\label{ps}
\end{eqnarray}
Since $M > M^*$ due to the Lorentz scalar potential, 
we have always $\delta \langle r_p^2 \rangle > 0$ in the relativistic 
model. In the present calculation,
we have determined the coupling constants so as to reproduce the
nucleon binding energy and nuclear density of nuclear matter.
Then, we obtain 
\begin{eqnarray}
\left(\frac{\langle r_p^2 \rangle^* }{\langle r_p^2 
\rangle}\right)^{1/2} = 1.146, \ \ \ \ \ \ ( M^* = 0.731 M ).
\label{rad}
\end{eqnarray}
Thus, since the effective nucleon mass is smaller in the nuclear
medium, the proton size is increased. This means that the nucleon
form factor is reduced. This is the reason why the Coulomb sum
values are strongly quenched due to the anti-nucleon degrees of freedom.
We note that if we define the effective $\omega$-meson mass $m^*_{\rm v}$ by
its RPA self-energy, Eq.(\ref{ps}) is expressed as\cite{k6}
\begin {eqnarray}
\delta \langle r_p^2 \rangle \approx 3\left( \frac{1}{m^{*2}_{\rm v}} - 
\frac{1}{m^2_{\rm v}}\right),
\end{eqnarray}
as expected in the vector-meson dominance model. The quenching
of the Coulomb sum is also interpreted as the reduction of
the $\omega$-meson mass in nuclear medium, 
which is $m^*_{\rm v} = 0.696 m_{\rm v}$ 
in the present model.

In Fig. 2 experimental data are shown from Saclay\cite{m1} and SLAC\cite{che}
which seem 
to support the quenching, but there are discussions on their
analyses of the data\cite{j,m2}. The prediction of the relativistic
model should be examined in more detail with new experiment.

Finally we should mention that the modified nucleon form factor
slightly improves also an agreement of elastic-scattering cross section
with experiment. In Fig. 4 is shown the result of NL-SH for the nucleon form
factor corresponding to the radius Eq.(\ref{rad})
with the same designation  as in Fig.1. 

\section{Conclusions}

The relativistic and non-relativistic model work well
for low-energy nuclear phenomena, but explain them
in different ways from each other.
In high momentum transfer phenomena,
however, there is a difference between their predictions.
The relativistic effects in the relativistic model
are not obtained as a simple correction to the non-relativistic
models.

First we have shown that the relativistic model reproduces
the difference between the cross sections for elastic electron
scattering from $^{40}$Ca and $^{48}$Ca, which has not been
explained in non-relativistic models. The time-component
of the relativistic four-current includes the neutron spin-orbit
current which is enhanced by the Lorentz scalar potential.
Electron scattering off unstable nuclei 
may be useful for distinguishing the relativistic from the
non-relativistic model in more detail. 

Second, we have shown a role of 
the anti-nucleon degrees of freedom.
 They are required from
the fundamental reason in the relativistic model,
but they are hidden in low-energy phenomena.
Within the same framework for the 
low-energy phenomena, however, it is shown that effects of 
the anti-nucleons with the effective mass 
appear in the Coulomb sum rule at high-momentum transfer.
Since the anti-nucleon degrees of freedom are an important ingredient
of the relativistic model, observation of those effects is essential
for the question if the relativistic model is realistic. 
New experiment
on the Coulomb sum values around the momentum transfer 1 GeV is
desirable in order to answer this question. 
It would be very serious for 
the relativistic model, if the Coulomb sum values are not quenched,
compared with the non-relativistic one.

\renewcommand{\baselinestretch}{1}\large

\newpage

\begin{figure}[h]
   \begin{center}
   \includegraphics[height=22pc]{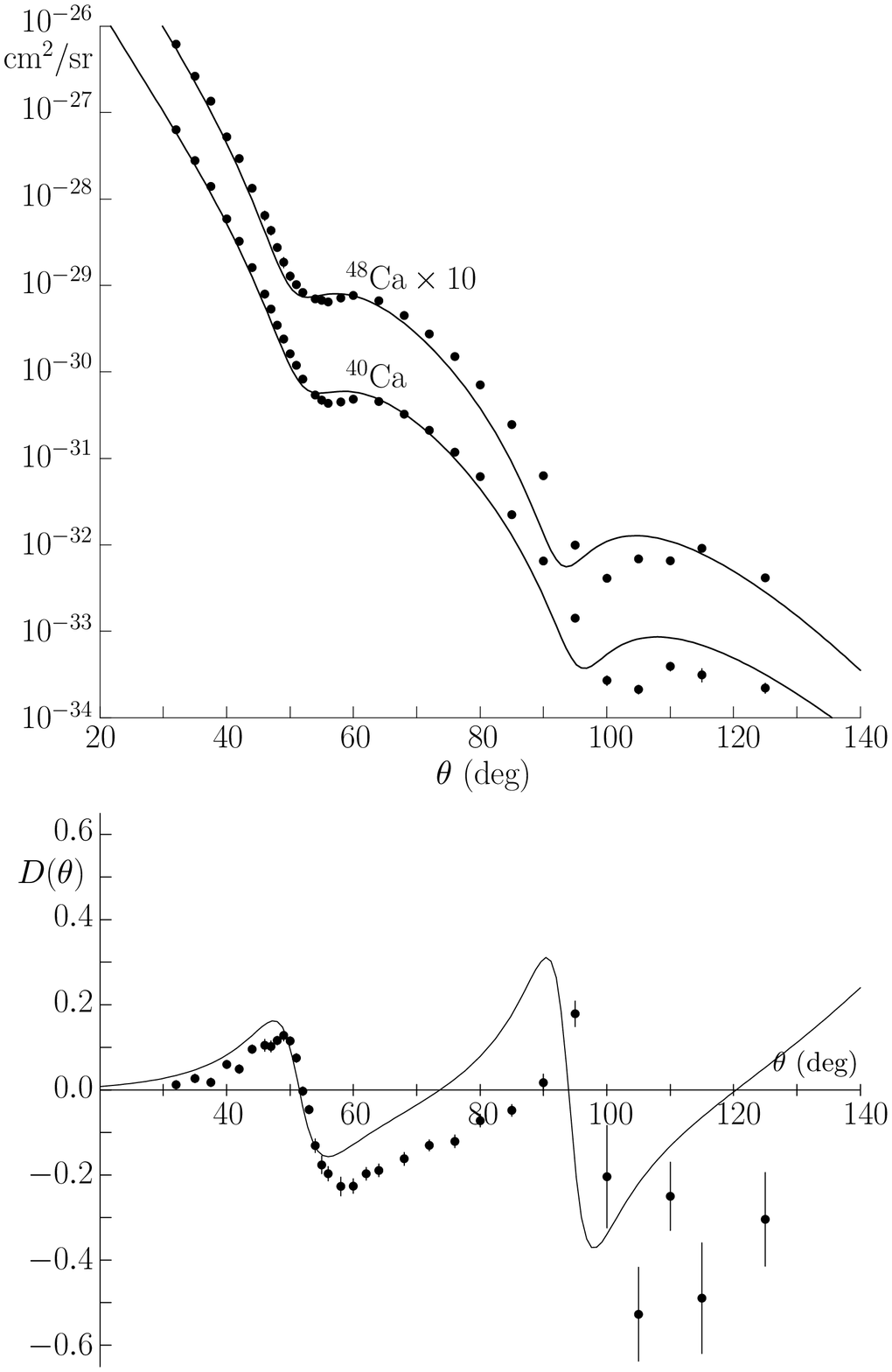}
\qquad   \includegraphics[height=22pc]{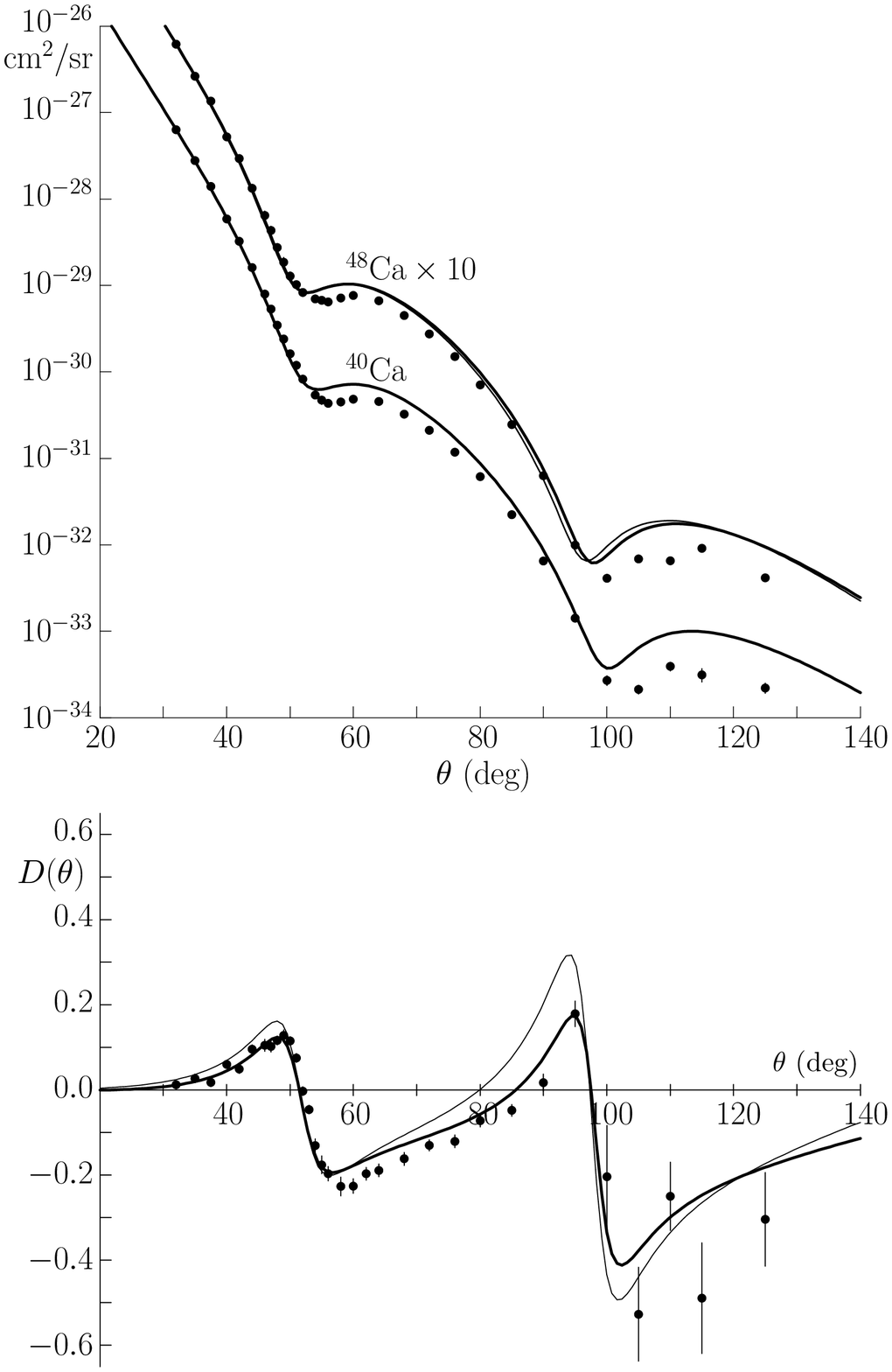}
   \end{center}
   \caption{Elastic-scattering cross sections for $^{40}$Ca and 
    $^{48}$Ca and their difference. For the details, see the text.} 
\end{figure}

\newpage

\begin{figure}[h]
   \begin{center}
   \includegraphics[height=24pc]{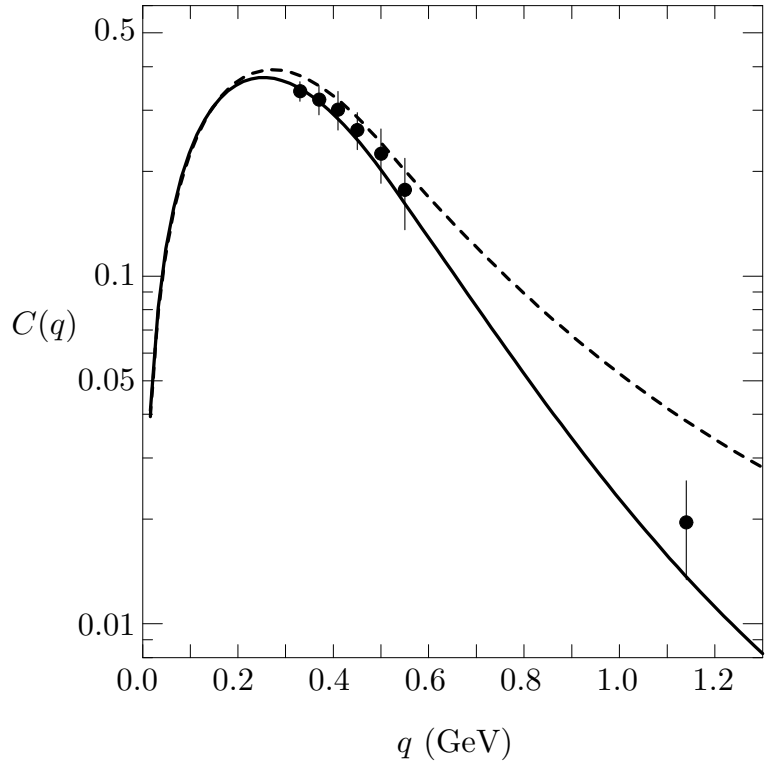}
   \end{center}
   \caption{The Coulomb sum values as a function of the momentum transfer. 
   For the details, see the text.} 
\end{figure}

\newpage

\begin{figure}[h]
   \begin{center}
   \includegraphics[height=6pc]{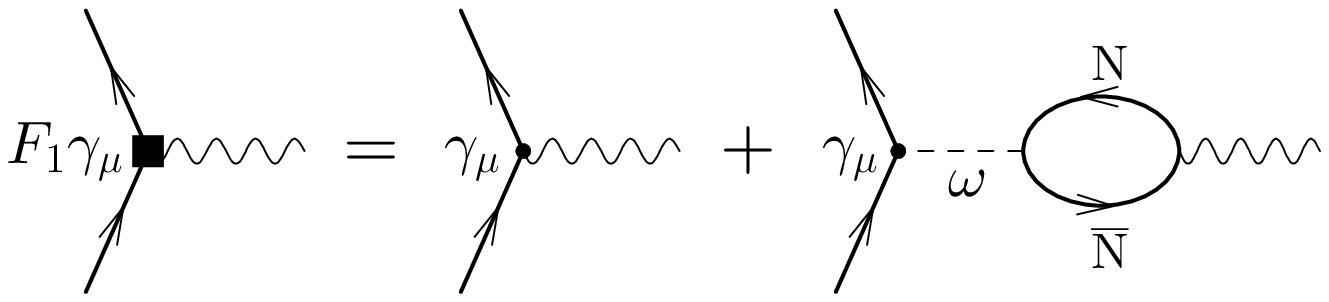}
   \end{center}
   \caption{Dirac form factor.} 
\end{figure}

\newpage

\begin{figure}[h]
   \begin{center}
   \includegraphics[height=30pc]{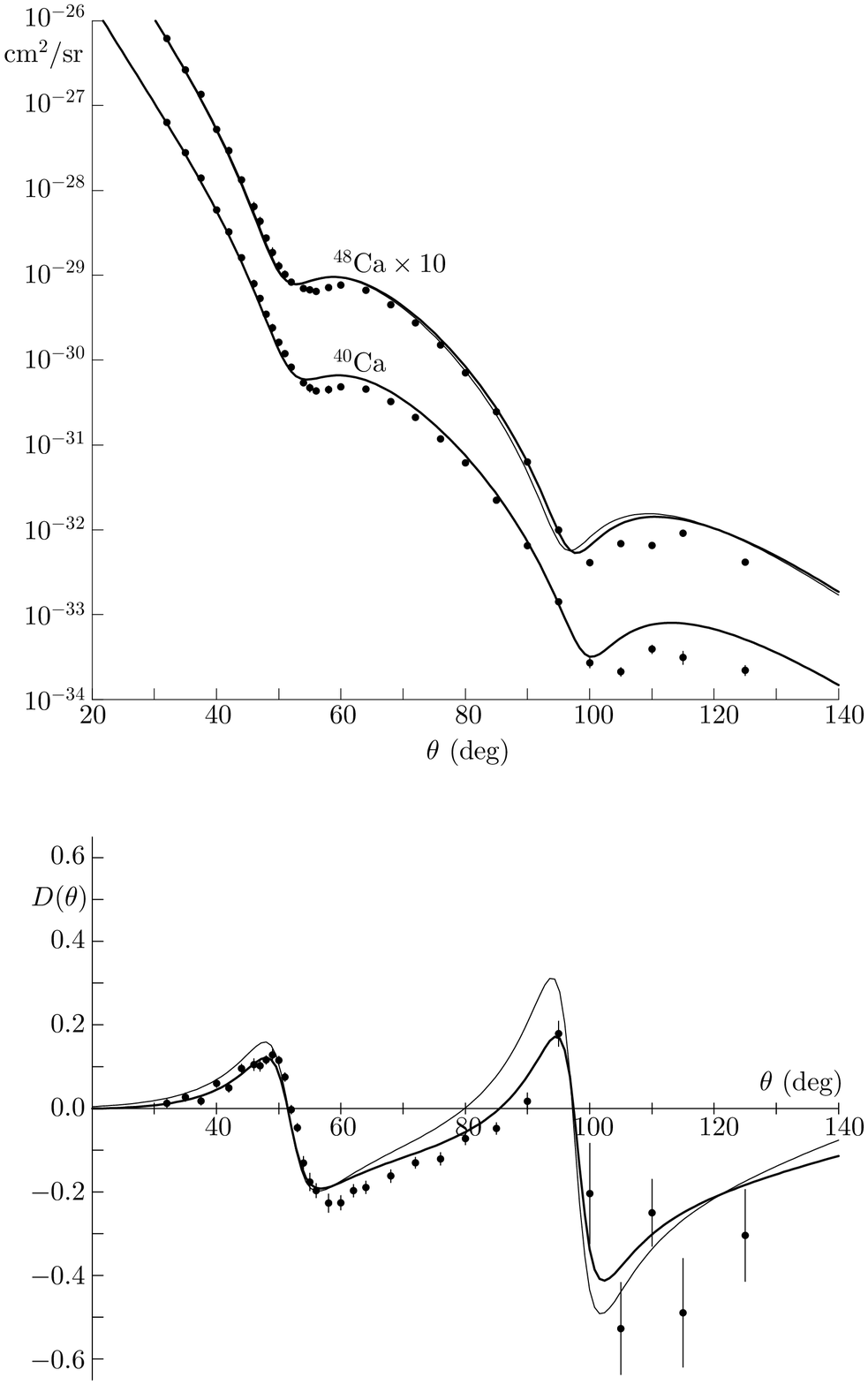}
   \end{center}
   \caption{Same as Fig. 1, for NL-SH with the nucleon form factor corresponding to the radius Eq.(\ref{rad})} 
\end{figure}

\end{document}